\documentclass[aps,preprint]{revtex4}
\usepackage{graphicx,amssymb}
\begin{document}
\setcounter{page}{1}

\title{ Upper Critical Field   Based on the Width of  $\Delta$H = $\Delta$B region\\ in a Superconductor}

\author{H. B. \surname{Lee}}
\author{G. C. \surname{Kim}}
\author{B. J. \surname{Kim}}
\author{Y. C. \surname{Kim}}
\email{yckim@pusan.ac.kr}
\thanks{Fax: +82-51-513-7664}
\affiliation{Department of Physics, Pusan National University, Busan 46241, Korea}

\begin{abstract}

We studied a method of measuring upper critical field (H$_{c2}$) of a superconductor based on the width of $\Delta$H = $\Delta$B region, which appears in the superconductor that volume defects are  many and dominant. Here we present the basic concept and details of the method. 
Although H$_{c2}$ of a superconductor is fixed according to kind of the superconductor, it is difficult to measure H$_{c2}$  experimentally, and the results are different depending on the experimental conditions. 
H$_{c2}$ was calculated from the theory that pinned fluxes at  volume defects are picked out and move into an inside of the superconductor when  their arrangement  is the same as that of H$_{c2}$ state of the superconductor.  H$_{c2}$ of MgB$_2$ obtained by the method was 65.4 Tesla at 0 K. The reason that H$_{c2}$ obtained by the method is closer to ultimate H$_{c2}$ is based on that $\Delta$F$_{pinning}$/$\Delta$F$_{pickout}$ is more than 4 when  pinned fluxes at  volume defects of 163 nm radius are picked out. 
The method will help to find the ultimate H$_{c2}$ of volume defect-dominating superconductors.

 \end{abstract}

\pacs{74.60.-w; 74.70.Ad}

\keywords{ MgB$_2$, Flux pinning effect, FeTi particles}

\maketitle

 Substantially, upper critical field (H$_{c2}$)  is difficult to measure at 0 K in a type II superconductor. 
It is because H$_{c2}$ of the type II superconductor is very high besides the difficulty of temperature. When a high magnetic field is applied, most of the fields penetrate into the inside of the superconductor, which result in extremely low diamagnetic property by 4$\pi$M = B - H, where M , B and H are magnetization, magnetic induction and applied magnetic field, respectively. 

Nevertheless, knowing the ultimate H$_{c2}$ of a superconductor  is  important, which is defined as  H$_{c2}$ in the ideal state. Practically, it is important to know how high the superconductor maintains its superconductivity in magnetic field, and more important thing is that H$_{c2}$ of a superconductor tell us one of  important parameters of the superconductor, which is coherence length ($\xi$) ($\Phi_o$/2$\pi$$\xi^2$ = H$_{c2}$, $\Phi_o$ is flux quantum which is 2.07$\times10^{-7}$ G$\cdot$cm$^2$) \cite{Poole}.

There are roughly  three methods to obtain H$_{c2}$  of a superconductor. The first is  the method of flowing currents after applying a high magnetic field and checking voltages (currents method). Many researchers are using this method, and the results  are  relatively reliable \cite{Kijoon, Flukiger}. However, there is much controversy for which point between the onset and the offset should be regarded as H$_{c2}$, and result voltages are also influenced by current density \cite{Hempstead}. 
In addition, in order to make results reliable, measured value should be al least obtained at 5 K. A great deal of efforts to set the equipment are required because H$_{c2}$ at 5 K is usually not small. 
Nonetheless, Gurevich et al. used this method  measuring H$_{c2}$ of MgB$_2$ film and reported that H$_{c2}$ is approximately 48 Tesla (T)  at 0 K \cite{Gurevich}. 

The second method is to measure critical temperature (T$_c$) and to determine H$_{c2}$ by theory (H$_{c2}$=1.83T$_c$) \cite{Poole1}. Theoretically, H$_{c2}$ of MgB$_2$ is approximately 68.6 T at 0 K when critical temperature (T$_c$) is 37.5 K. The third is to extrapolate H$_{c2}$ to 0 K after H$_{c2}$ were measured at various temperatures. The extrapolation  uses the property that H$_{c2}$  of a superconductor increases when the temperature decrease (M-H curve method). Of course, this method is convenient, but there is a problem that a limit of the magnetic field exists in equipment and  the result are hard to be believed if applied magnetic field does not increases  carefully in high magnetic field region.  Therefore, the reliability is low.

Since a diamagnetic property is extremely small if applied magnetic field is high, superconductor becomes very vulnerable to external influences.  The behaviors do not differ at 15 K and 20 K because H$_{c2}$ of the temperatures are considerably large in MgB$_2$.  In addition, it may be considered that  H$_{c2}$  vary depending on the specimen in the M-H curve method as shown in the Fig. \ref{fig1}  because a diamagnetic property  varies greatly depending on the pinning state of  defects in relatively high field (6.5 T).  Nonetheless, it is certain that  H$_{c2}$  does not change. Generally,  it  was reported that H$_{c2}$ of MgB$_2$   is 20 - 30 T \cite{Buzea,  sung, Sologubenko}. 
 However, this is equivalent to the statement that  H$_{c2}$ of MgB$_2$ is more than 20 T and  the upper limit is unknown.

 No matter how high H$_{c2}$ was measured, it was the H$_{c2}$ that was appropriate for the condition of measurement and the state of the specimen, thus it has its own meaning. However, it is also important to  measure ultimate H$_{c2}$ experimentally.
In previous study,  we have asserted  that a $\Delta$H = $\Delta$B region is formulated  if volume defects are many enough  in volume defect-dominating superconductor, which is the region that increased applied magnetic field   is the same as increasing magnetic induction \cite{Lee4}.  Two conditions are suggested for that pinned fluxes have to be picked out from the volume defect, which are $\Delta$F$_{pinning}$ $<$ $\Delta$F$_{pickout}$ or the arrangement of the pinned fluxes at a volume defect is equal to that of H$_{c2}$. We have calculated H$_{c2}$  using the theory with experimental results, and have obtained a fairly reasonable result. Thus, we would introduce the method of obtaining H$_{c2}$ of superconductor  based on a $\Delta$H = $\Delta$B region.

Pure MgB$_{2}$ and (Fe, Ti) particle-doped MgB$_{2}$ specimens were synthesized using the nonspecial atmosphere synthesis (NAS) method \cite{Lee}. Briefly,
NAS method needs Mg (99.9\% powder), B (96.6\% amorphous powder), (Fe, Ti) particles and stainless steel tube. 
  Mixed Mg and B stoichiometry, and  (Fe, Ti) particles were added by weight. They were finely ground and pressed into 10 mm diameter pellets. (Fe, Ti) particles were ball-milled for several days, and average radius of (Fe, Ti) particles was approximately 0.163 $\mu$m \cite{Lee4}. 
   On the other hand, an 8 m-long stainless-steel (304) tube was cut into 10 cm pieces. Insert holed Fe plate into stainless- steel (304) tube. One side of the 10 cm-long tube was forged and welded. The pellets and pelletized excess Mg  were placed at uplayer and downlayer in the stainless-steel tube, respectively. The pellets were annealed at 300 $^o$C  for 1 hour to make them hard before inserting them into the stainless-steel tube. The other side of the stainless-steel tube was also forged. High-purity Ar gas was put into the stainless-steel tube, and which was then welded. Specimens had been synthesized at 920 $^o$C  for 1 hour. They are cooled in air and quenched in water respectively. The field and temperature dependence of magnetization were measured using a MPMS-7 (Quantum Design). 

  Figure \ref{fig1} shows field dependences of magnetization (M-H curves)  for various  temperature, which is the maximum at 6.5 Tesla (T). They are not considered below 15 K because the diamagnetic property at the temperatures is quite meaningful at 6.5 T. The field that diamagnetic property changes from - to + is H$_{c2}$, and results are shown in Fig. \ref{fig2}. The equation of extrapolating is as follows.
\begin{eqnarray}
\xi(T)^2 
\propto\frac{1}{1-t}\Rightarrow
H_{c2}=\frac{\Phi_o}{2\pi\xi^2} \propto1-t
 \end{eqnarray} 
 where t = $T_m$/$T_c$, $\xi$ is  coherence length \cite{Poole, Tinkham}. $T_m$ is the measuring temperature  and  $T_c$ is critical temperature. 

As shown in Fig. \ref{fig2}, different results  were obtained for three specimens, and it needs to check that results were closer to the ultimate H$_{c2}$ of the specimen.  Since the diamagnetic properties of the superconductor approach zero if the external magnetic field is high enough, the arrangements of the flux quanta in the superconductor would be that of Fig. \ref{fig3}. When many magnetic fluxes quanta have penetrate into the inside of the superconductor, the supercurrent circulating the surface of the superconductor is  the  only one related to diamagnetic property. Thus, the field that superconducting current disappears would be H$_{c2}$ as the external field increases.

As shown in Fig. 1, applied magnetic field increased by 0.5 T when the external field is above 1 T. It is clear that all of increased magnetic field penetrates the inside of the superconductor when the diamagnetic property is close to zero. On the other hand, the fluxes in the superconductor exist as flux quanta, and the repulsive forces are acting between them. The repulsive force per unit length (cm) between quantum fluxes is, which is caused by vortexes
{\setlength\arraycolsep{2pt}
\begin{eqnarray}
f = J_s\times\frac{\Phi_o}{c}    
\end{eqnarray} 
where $J_s$ is the total supercurrent  density due to vortices  \cite {Tinkham2}.

When a large field of 0.5 T increases at once, the flux quanta that try to penetrate into the superconductor will interact with the existing flux quanta  in the superconductor, and the repulsive force between them will cause continuous vibration. The vibration will cause mutual interference, which are amplification and attenuation. The attenuation is expected to have little effect on the diamagnetic supercurrent, but the amplification may cause some flux quanta to rebound out from the superconductor because there is little difference in fluxes density between the inside and the outside of the superconductor and circulating supercurrent is tiny.

When rebounds of quantum fluxes occur, the supercurrents circulating the surface of the superconductor are interfered.  When the number of the rebounding flux quantum exceeds a certain value, the supercurrents disappear, which results in that the diamagnetic property does not appear. It is considered that the phenomenon increases as the distance between fluxes  become closer to that of ultimate H$_{c2}$, and increases as the magnitude of magnetic field applied at once increases. Therefore, it is clear that H$_{c2}$ obtained from M-H curves must be  much lower than the ultimate H$_{c2}$.

  On the other hand,  when the magnetic field is  applied first to the superconductor and next the superconducting current is supplied to determine H$_{c2}$ of a superconductor, fluxes inside the superconductor would be much more stabilized.  In this case, it is determined that the magnetic field in the superconductor is arranged in the form of Fig. 3. (a) or ultimately stabilized form of Fig. 3. (b). From the point of view, it is considered that the arrangement of the flux quanta in the H$_{c2}$ state of a superconductor is not depending on what kind of superconductor is, but how long it takes after applying the magnetic field \cite{Abrikosov,  Huebener}. Since a stabilization of quantum fluxes is achieved over time, it is natural that the arrangement of the flux quanta will be the  state as shown in Fig. 3. (b). 

If a small amount of current  flows around  the surface of specimen in the state that the magnetic fluxes in the superconductor are stabilized, the increasing magnetic  flux quanta in the superconductor is only a magnetic field generated by the flowing current. If the magnitude of the current is small, the interference caused by the increased magnetic flux quanta would be small, thus the magnetic flux quanta rebounding out of the superconductor will also be small. Therefore, it is considered that H$_{c2}$ measured by currents method is much closer to the ultimate H$_{c2}$ of the superconductor than that of M-H curves. However, it is certain that H$_{c2}$ measured by this method is not the ultimate because a certain amount of current must flow, which generates a magnetic field . 

If volume defects are spherical, their size is constant, and they are arranged regularly in a superconductor, a superconductor of  1 cm$^3$ has m$'^3$ volume defects. Assuming that the pinned fluxes at volume defects are peaked out and move into an inside of the superconductor when the arrangement of pinned fluxes is the same as that of H$_{c2}$ as shown in Fig. \ref{fig3} (a), the maximum number of flux quanta that can be pinned at a spherical defect of radius $r$ 
 in a static state is
\begin{eqnarray}
n^2 = \frac{\pi r^2}{\pi (\frac{d}{2})^2}\times P = (\frac{2r}{d})^2\times P = \frac{\pi r^2}{d^2}
 \end{eqnarray} 
 where $r$, $d$ and $P$ is the radius of defects, the distance between quantum fluxes pinned at the volume defect of which radius r is and filling rate which is $\pi$/4 when they have square structure, respectively, as shown in Fig. \ref{fig3} (a) \cite{Lee5}.

If volume defects in a superconductor  are many enough, the superconductor has a $\Delta$H = $\Delta$B region, and the width of the region is
\begin{eqnarray}
W_{\Delta H=\Delta B} = H_{final} - H_{c1}'=n^2m_{cps}m\Phi_0 -4\pi M-H_{c1}'
 \end{eqnarray} 
where  $H$$_{final}$ is the final field of  the $\Delta$H = $\Delta$B region, $H_{c1}'$ is the first field of  the $\Delta$H = $\Delta$B region  \cite{Lee4}. 
$m$$_{cps}$ is the number
of defects which are in the vertically closed packed state, $n$$^2$ is the number of flux quanta pinned at a defect of radius r,  $m$ is the number of the volume defects from surface to center along an axis ($m'$=2$m$),  $M$ is magnetization, and $\Phi_0$  is flux quantum. 
$m$$_{cps}$ is the minimum number of defects when the penetrated fluxes into the superconductor are completely pinned. Thus, 2$r$$\times$$m_{cps}$ is unit.


The number of flux quanta pinned at a defect is 
\begin{eqnarray}
n^2= \frac{  W_{\Delta H=\Delta B} + 4\pi M +  H_{c1}'}{m_{cps}m\Phi_0 }
 \end{eqnarray} 
Arranging after the equation is put into Eq. (3),
\begin{eqnarray}
d^2 =  \frac{\pi r^2 m_{cps}m\Phi_0}{ W_{\Delta H=\Delta B} + 4\pi M +  H_{c1}'}
 \end{eqnarray} 
Therefore
\begin{eqnarray}
H_{c2} = \frac{\Phi_0}{d^2}
=\frac{( W_{\Delta H=\Delta B} + 4\pi M +  H_{c1}')}{\pi r^2 m_{cps}m}
\end{eqnarray}
Because 2$r$$\times$$m_{cps}$ = 1, the equation is
\begin{eqnarray}
H_{c2} = \frac{\Phi_0}{d^2}
=\frac{2(W_{\Delta H=\Delta B} + 4\pi M +  H_{c1}')}{\pi r m}
\end{eqnarray}

The thickness of the specimen used for the measurement is 0.25 cm. Thus, the width of the region as unit length have to be 4$\times$t (t is the thickness of the specimen). In addition, since applied magnetic field penetrates both sides of the specimen,  volume defects inside the superconductor have pinned the fluxes for both side until applied magnetic field reach H$_{c1}'$. 
Therefore, the width of the region as unit length is
\begin{eqnarray}
 W_{\Delta H=\Delta B} = n_d w + (n_d-1)H_{c1}' 
 \end{eqnarray}  
where 
$w$ is experimentally obtained width of the region, $n_d$ is the number of specimen when a specimen of unit length was divided ($n_d t$=1). 
Although the width of the region is 1.3 T as shown in the Fig. \ref{fig4} (b), the width of $\Delta$H = $\Delta$B region as unit length is 5.8 T because the width of the specimen was 0.25 cm.  

If the average radius of  defects  is 163 nm, the width of $\Delta$H = $\Delta$B region is 5.8 T, $M$ is 150 emu/cm$^3$, H$_{c1}'$ is 2000 Oe, and $m$ is 4000, which are experimental results of 5 wt.\% (Fe, Ti) particle-doped MgB$_2$ as shown in Fig. \ref{fig4},  H$_{c2}$ of the specimen is 56.7 T at 5 K. Concerning $m$, although the specimen have 8000$^3$ volume defects of 163 nm radius, it is 4000 because  magnetic field penetrates into the superconductor from both sides  \cite{Lee4}. 
The coherence length ($\xi$) is 2.41 nm when H$_{c2}$ is 56.7 T at 5 K. If  extrapolated  by Eq.(1), $\xi$ is 2.24 nm and H$_{c2}$ is 65.4 T at 0 K.

 As mentioned earlier, the methods of measuring H$_{c2}$ of a superconductor have their own drawbacks. Supercurrents method may be close to the ultimate H$_{c2}$ of the superconductor, but it is clear that there is a difference between the result and the ultimate H$_{c2}$ because of the magnetic field induced by applied currents. However, we believe that H$_{c2}$ measured by this method can further reduce  the difference.

 We could understand how stabilized the pinned fluxes are  in  H$_{c2}$ state if inspecting the force balances of the pinned fluxes when they are picked out. Generally, pinned fluxes at volume defect move when $\Delta$F$_{pickout}$ is more than $\Delta$F$_{pininng}$. However, it was our assertion that the pinned fluxes are picked out and moved even in  $\Delta$F$_{pinning}$$>$$\Delta$F$_{pickout}$ state when the flux arrangement is equal to that of H$_{c2}$. The justification of the assumption is that there is no pinning effect if the  neighborhoods of the volume defect are changed to normal state.

$\Delta$F$_{pinning}$ is 
\begin{eqnarray} 
\Delta F_{pinning}=\frac{\partial G}{\partial r} = - \frac{H_{c1}'^2}{8\pi}\times 4\pi r^2 + \frac{2n^2\Phi_o^2}{8\pi} 
\end{eqnarray}
and $\Delta$F$_{pickout}$   is
\begin{eqnarray}
\Delta F_{pickout} = \frac{aLA\Phi_o}{\pi c r^2}n^5
\end{eqnarray}
where $A$ is 0.103 A (Ampere), $aL$ is a average length of  quantum fluxes which are pinned and bent between defects ($a$ is average bent constant, $1<a<1.2$) and $c$ is the  velocity of light \cite{Lee5}.

Numerically, If H$_{c1}'$ is 2000 Oe, $r$ is 0.163 nm, $n$ is 45, and  $aL$ is  1.1$\times$3.9$\times$10$^{-4}$ cm, which are results of idealized 5 wt.\% (Fe, Ti) doped MgB$_2$ specimen, $\Delta$F$_{pinning}$ is 5.3$\times$10$^{-4}$ dyne and $\Delta$F$_{pickout}$ is 1.4$\times$10$^{-4}$ dyne.  Comparing $\Delta$F$_{pinning}$ with $\Delta$F$_{pickout}$, 
 $\Delta$F$_{pinning}$ /$\Delta$F$_{pickout}$  is more than 4. 
Generally, when fluxes are approaching a volume defect, they have a velocity.  If $\Delta$F$_{pinning}$  are similar with $\Delta$F$_{pickout}$, the depinning  of pinned fluxes from the volume defect is easier  than that of calculation because fluxes have a velocity when they move in the superconductor. However, if $\Delta$F$_{pinning}$ is more than 4 times of $\Delta$F$_{pickout}$, it is considered that the depinning occurs after the arrangement of the fluxes becomes like that of H$_{c2}$ even if fluxes had some velocity.

We have investigated characteristics of several methods of obtaining H$_{c2}$ of  type II superconductors and explained that any existing experimental method to obtain H$_{c2}$ would be different from the ultimate H$_{c2}$. In addition, no matter how high   H$_{c2}$ was obtained, it has its meaning because it was caused by the state of the specimen and measurement conditions.
 We suggested  a method to obtain  H$_{c2}$, which is that H$_{c2}$ of volume defect-dominating superconductor was obtained from the width of  $\Delta$H = $\Delta$B. We used the property that  $\Delta$H = $\Delta$B region formulated in the M-H curve  when the volume defects in the superconductor are many enough. It is based on the theory that the arrangement of the pinned fluxes at the volume defects will be picked out from the volume defects and move  when the arrangement of them equals to that of H$_{c2}$. From the results of 5 wt.\% (Fe, Ti) doped MgB$_2$, H$_{c2}$ was 56.7 T at 5 K. We obtained that $\Delta$F$_{pinning}$/$\Delta$F$_{pickout}$ is more than 4, that means that  fluxes had pinned at the volume defect depinned  even though $\Delta$F$_{pinning}$ is much larger than $\Delta$F$_{pickout}$.  The behavior result in that the H$_{c2}$ might be less sensitive to fluctuation. Therefore, it is determined that the obtained H$_{c2}$ by the method  is much closer to the ultimate H$_{c2}$ of the superconductor.

\vspace{2cm}


\newpage

\begin{figure}
\vspace{2cm}
\begin{center}
\includegraphics*[width=17cm]{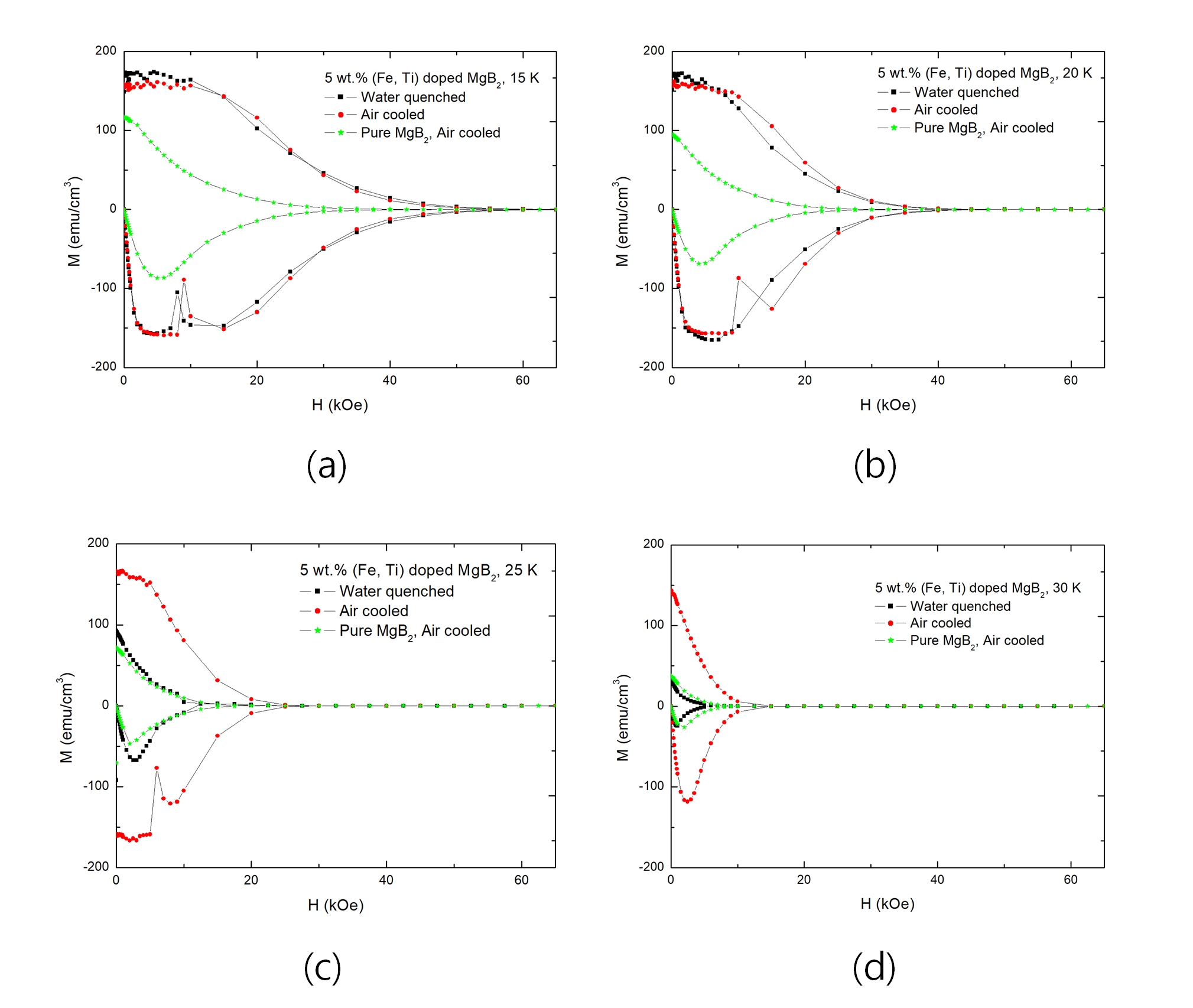}
\end{center}
\caption{ Field dependences of magnetization (M-H curves) of pure MgB$_2$ and 5 wt.\% (Fe, Ti) doped MgB$_2$ at various temperatures. Pure MgB$_2$ was air-cooled. 5 wt.\% (Fe, Ti) doped MgB$_2$ were  air-cooled and water-quenched, respectively. (a): M-H curves at 15 K. (b): M-H curves at 20 K. (c): M-H curves at 25 K. (d): M-H curves at 30 K}
\label{fig1}
\end{figure}

\begin{figure}
\vspace{2cm}
\begin{center}
\includegraphics*[width=15cm]{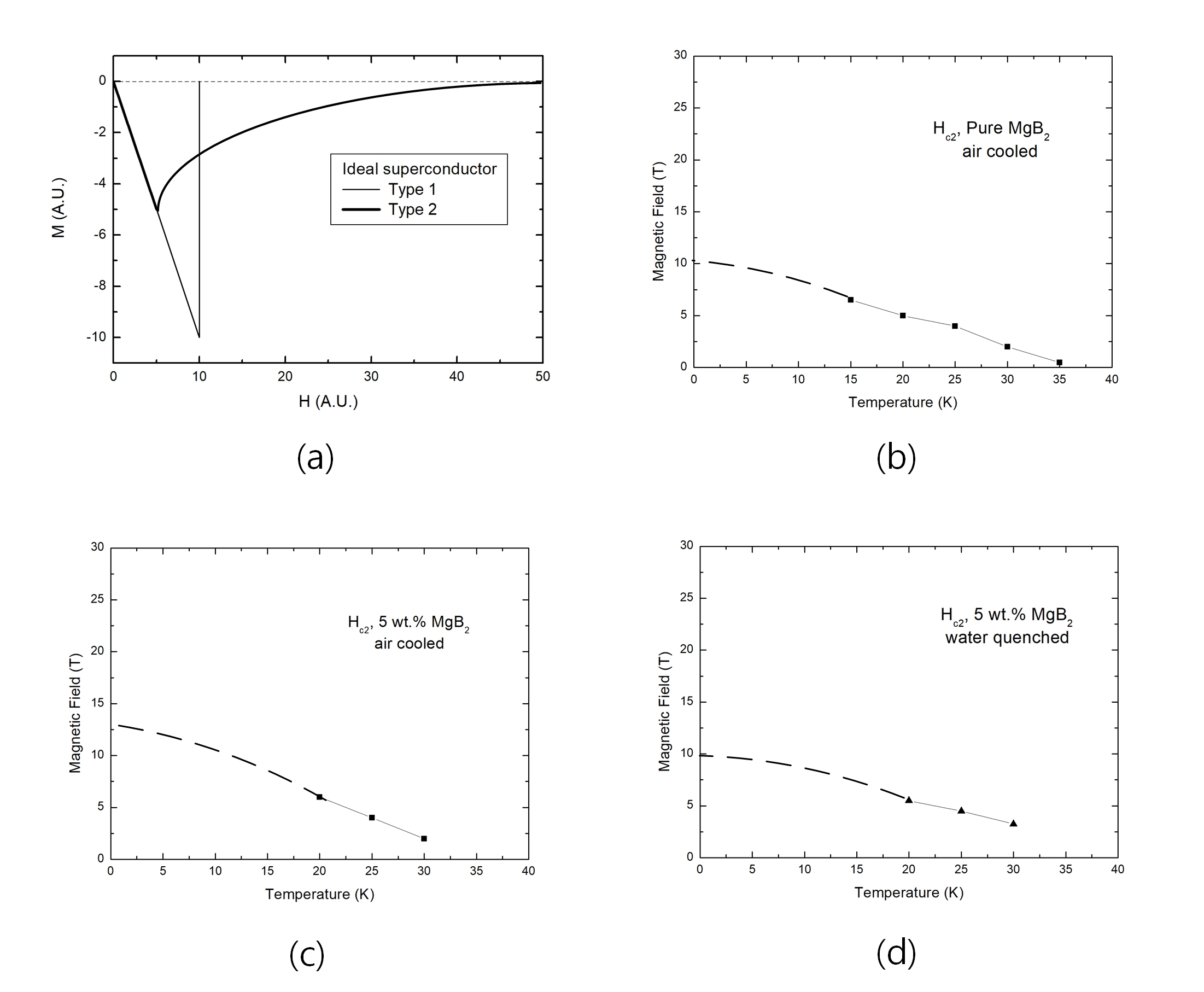}
\end{center}
\caption{ Field dependences of magnetization (M-H curve) for ideal superconductor  and extrapolations of  H$_{c2}$ for various specimens. (a): H$_{c1}$ and H$_{c2}$ of ideal superconductor. (b): H$_{c2}$ of pure MgB$_2$, which was air-cooled. (c) H$_{c2}$ of 5 wt.\% (Fe, Ti) doped MgB$_2$, which was air-cooled. (d): H$_{c2}$ of 5 wt.\% (Fe, Ti) doped MgB$_2$, which was water-quenched.
 }
\label{fig2}
\end{figure}

\begin{figure}
\vspace{2cm}
\begin{center}
\includegraphics*[width=11cm]{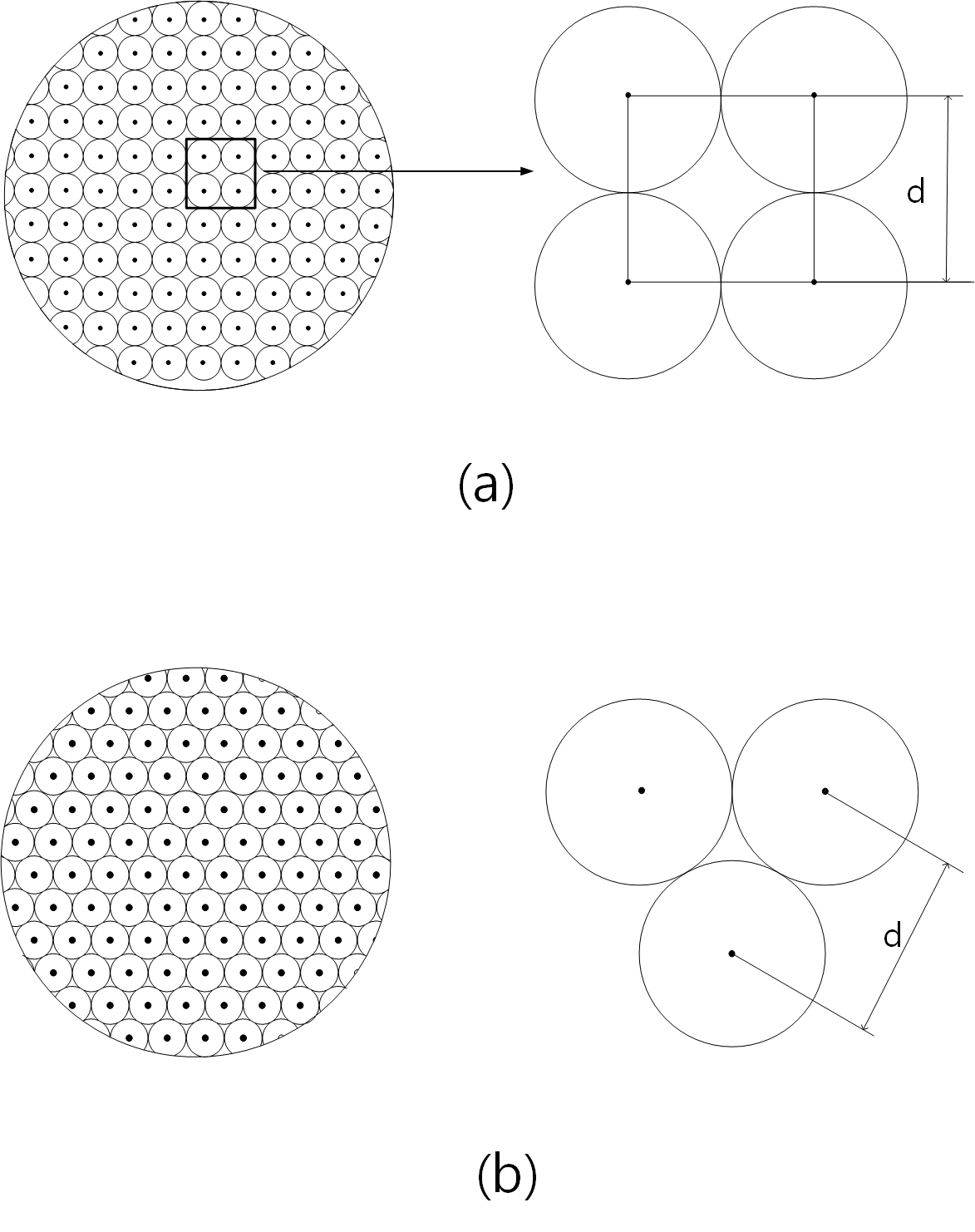}
\end{center}
\caption{Schematic representations for an arrangement of flux quanta at H$_{c2}$. (a): Square form of flux quanta at H$_{c2}$. (b): Triangular form of flux quanta at H$_{c2}$.
 }
\label{fig3}
\end{figure}

\begin{figure}
\vspace{2cm}
\begin{center}
\includegraphics*[width=10cm]{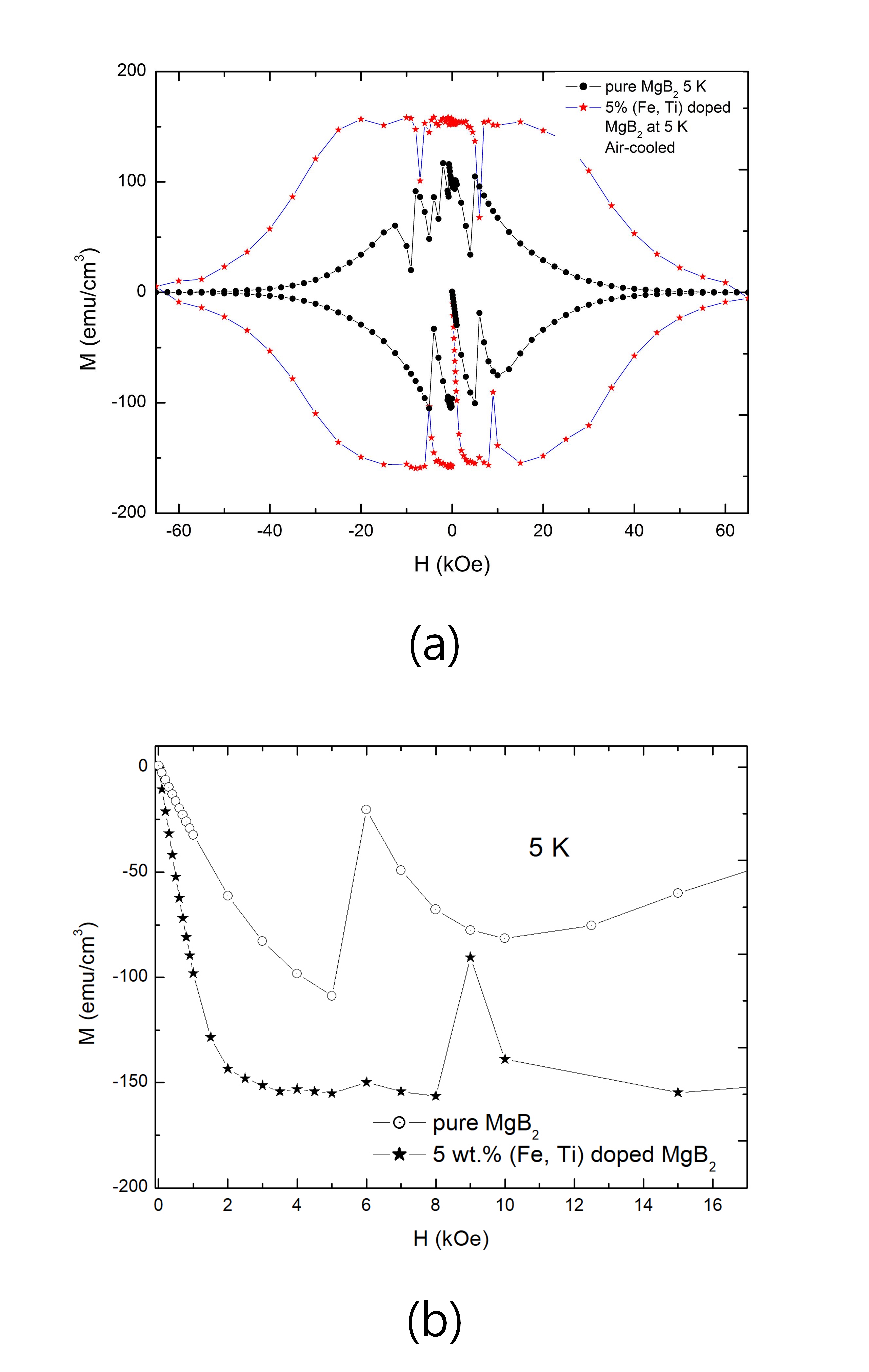}
\end{center}
\caption{A decision of the width of the $\Delta$H = $\Delta$B region from field dependences of magnetization (M-H curves) for pure MgB$_2$ and  5 wt.\% (Fe, Ti) doped MgB$_2$. (a): Full M-H curves. (b): Zero width of the $\Delta$H = $\Delta$B region in pure MgB$_2$ and 1.3 T (1.5 T - 0.2 T)  of $\Delta$H = $\Delta$B region in 5 wt.\% (Fe, Ti) doped MgB$_2$, which was air-cooled.}
\label{fig4}
\end{figure}

\end{document}